\documentclass[10pt]{elsart}

\newcommand{\cal}[1]{{\mathit #1}}
\newcommand{\simgeq}{\; \raisebox{-0.4ex}{\tiny$\stackrel
{{\textstyle>}}{\sim}$}\;}
\newcommand{\simleq}{\; \raisebox{-0.4ex}{\tiny$\stackrel
{{\textstyle<}}{\sim}$}\;}
\newcommand{\up}[2]{$^{#1#2}$}

\newcommand{\BibTitle}[1]{#1}
\renewcommand{\BibTitle}[1]{}
\newcommand{\doetal}[1]{\ {\em et al\/}}

\newcommand{\refeq}[1]{(\ref{#1})}

\usepackage{psfig}
\usepackage{bm}

\begin{document}
\small

\begin{frontmatter}
\title{Neutral-current neutrino-nucleus cross-sections for
  $A\sim50-65$ nuclei}

\author[ORNL,UTK]{A.\ Juodagalvis\thanksref{eaddr}},
\author[Arhus]{K.\ Langanke},
\author[Spain1,Spain2]{G.\ Mart{\'\i}nez-Pinedo},
\author[ORNL,UTK]{W.R.\ Hix},
\author[ORNL]{D.J.\ Dean},
and
\author[Arhus]{J.M.\ Sampaio}

\address[ORNL]{Physics Division, Oak Ridge National Laboratory, P.O.\
  Box 2008, Oak Ridge, TN 37831-6373, USA}
\address[UTK]{The Department of Physics and Astronomy, University of
  Tennessee, Knoxville, TN 37996, USA}
\address[Arhus]{Institut for Fysik og Astronomi, {\AA}rhus
  Universitet, DK-8000 {\AA}rhus C, Denmark}
\address[Spain1]{Institut d'Estudis Espacials de Catalunya, Edifici
  Nexus, Gran Capit{\`a} 2, E-08034 Barcelona, Spain}
\address[Spain2]{Instituci{\'o} Catalana de Recerca i Estudis Avan{\c
  c}ats, Llu{\'\i}s Companys 23, E-08010 Barcelona, Spain}

\thanks[eaddr]{e-mail: andrius@mail.phy.ornl.gov}

\received{23 April 2004}


\begin{abstract} 
 We study neutral current neutrino-nucleus reaction
 cross-sections for Mn, Fe, Co and Ni isotopes. An
 earlier study for a few selected nuclei 
 has shown that in the supernova environment the
 cross sections are increased for low energy neutrinos due to
 finite-temperature effects. Our work
 supports this finding for a much larger set of nuclei.
 Furthermore we extend previous work to higher neutrino energies
 considering allowed and forbidden multipole contributions to the
 cross sections. The allowed contributions are derived from large-scale
 shell model calculations of the Gamow-Teller strength, while the
 other multipole contributions are calculated within the Random Phase
 Approximation.
 We present the cross sections as functions of initial and final
 neutrino energies and for a range of supernova-relevant temperatures.
These cross sections will allow improved
 estimates of inelastic neutrino reactions on nuclei
 in supernova simulations.
\end{abstract}

\begin{keyword}
Shell model; Random-Phase Approximation; Gamow-Teller transitions;
neutral-current; weak-interactions; 
neutrino-nucleus reaction cross section; nuclear
astrophysics.
\PACS{26.50.+x; 23.40.Bw; 21.60.Cs; 97.60.Bw}
\end{keyword}
\end{frontmatter}

\section{Introduction}
\label{sect-Introduction}

Neutrino-nucleus reactions play essential roles in many astrophysical
applications. These include core-collapse supernovae
\cite{BetheReview}, explosive \cite{Hix03JPG} and r-process
nucleosynthesis \cite{rmp}, and observation of solar and supernova
neutrinos by terrestrial detectors \cite{Kolbe03}.

Stars with masses $\simgeq 10 M_\odot$ end their lives as
core-collapse supernovae. In the final stages of their evolution these
stars produce an inner core consisting of electrons and `iron nuclei'
(nuclei with mass numbers $A \sim 60$).  As these nuclei have the
highest binding energy per nucleon, 
additional fusion reactions 
cost, rather than generate, energy.  The stellar core has lost its
energy source, and once the core mass has grown larger than the
Chandrasekhar mass limit the star starts to collapse under its own
gravity.  The simulation of the final evolution and collapse of such
massive stars is one of the outstanding current challenges in
astrophysics, combining state-of-the-art microphysics input
(nuclear and neutrino physics) with sophisticated treatments of
hydrodynamics and radiation transport.  Despite significant progress
in both of these aspects, current core-collapse supernova simulations
often fail to yield explosions \cite{Mezzacappa01,Rampp00,Buras03}.
Since most of the explosion energy is carried by neutrinos, an
essential part of these simulations is a detailed treatment of
neutrino transport including the various interactions of neutrinos
with the supernova environment. One of these interaction processes is
inelastic neutrino-nucleus scattering, which is currently
ignored in supernova simulations. 
Recent work indicates that
this might be unjustified.
It has been shown that finite temperature effects increase the
low-energy neutrino-nucleus cross sections significantly
\cite{Sampaio02}.
Furthermore,
recent improvements in
calculating electron capture on heavy nuclei during the collapse phase
implies that these captures dominate the capture on free protons
\cite{Langanke03,Hix03}, in contrast to previous beliefs
\cite{BetheReview,Bruenn85}. 
Captures on nuclei produce neutrinos with smaller energies, making
temperature effects more important, and
the inclusion of inelastic neutrino-nucleus reactions in
supernova simulations more relevant.
A call for reliable neutrino-nucleus
cross sections has also been made in the context of explosive
nucleosynthesis, occurring when the shock wave passes through the
exploding star which leads to fast nuclear reactions \cite{Hix03JPG}.

The pioneering study of neutrino-nucleus reactions during 
core collapse
has been performed by Bruenn and Haxton \cite{BruennHaxton}.
These authors calculated inelastic neutrino-nucleus scattering rates
(neutral current) and neutrino-nucleus absorption rates
(charged current) approximating the nuclear composition present in the stellar
core by a single nucleus, $^{56}$Fe. The calculated rates were based on
a nuclear model appropriate for temperatures $T=0$, combining a
truncated nuclear shell model evaluation of the allowed contributions to
the cross sections with estimates for forbidden components within the
framework of the Goldhaber-Teller model \cite{BruennHaxton}. In detailed
supernova simulations Bruenn and Haxton studied 3 phases of the collapse
and explosion in which neutrino-nucleus reaction could be potentially 
important: the matter infall, the prompt shock propagation phase, and
the delayed-shock phase.
First, they found that inelastic
neutrino-nucleus scattering plays the `same extremely important role
of equilibrating $\nu_e$'s to matter during infall as neutrino-electron
scattering' \cite{BruennHaxton}. Second, they did not confirm an
earlier suggestion by Haxton \cite{Haxton88} who pointed to the
possibility that neutrino-nucleus reactions can preheat the matter ahead
of the shock during the early phase of the explosion. Third,
their simulation indicated no significant contribution
of neutrino-nucleus reactions to the revival of the stalled shock,
compared to neutrino absorption on free nucleons.

Sampaio {\em et al\/} have argued that the finite temperature environment
of a supernova alters inelastic neutrino-nucleus
scattering cross sections noticeably \cite{Sampaio02}.
Using
a shell-model treatment of the dominant Gamow-Teller (GT) component
they showed that finite-temperature effects increase the cross
sections for inelastic neutrino scattering on even-even nuclei, like
$^{56}$Fe, at low energies ($E_\nu\simleq 10$ MeV) significantly.  
On the other hand, their
study of $^{56,59}$Co and \up59Fe suggested that the effect is less
pronounced for odd-$A$ and odd-odd nuclei.  This variation in the
importance indicates that 
inelastic neutrino-nucleus
cross sections for application in supernova simulations should be
derived i) at finite temperatures and ii) for an appropriate matter
composition. 
The aim of the
present work is to prepare data on an ensemble of nuclei relevant for
supernova simulations.  
Calculations have been performed for 40 
isotopes of Mn,
Fe, Co, and Ni, covering in each isotope chain the range from
$N=Z$ to very neutron-rich nuclei. 
To estimate the possible effect of inelastic neutrino-nucleus
scattering within a supernova, we use the calculated cross sections
to  speculate about the matter heating rate in the post-bounce supernova.

In a supernova, inelastic neutrino-nucleus scattering occurs at rather
modest neutrino energies, $E_\nu \simleq 50$ MeV. This makes the
cross sections sensitive to nuclear structure effects, particularly
for neutrinos with energies lower than 20 MeV.  Therefore
nuclear models must be employed, which can describe the
many-body correlations in the nucleus accurately. The model of choice
is the shell model, which now allows for virtually converged
calculations of the Gamow-Teller response for nuclei in the iron mass
range \cite{Caurier99}, with a quite satisfactory reproduction of the
available experimental Gamow-Teller data
\cite{Caurier99,Frekers,Hagemann,Baeumer}. While the GT component
determines the neutrino-nucleus cross section at low neutrino
energies, higher multipole contributions become increasingly important
at higher neutrino energies. We have calculated these contributions
within the Random Phase Approximation (RPA)~\cite{Kolbe92}.

This paper is organized as follows. In section \ref{sect-Models} we
briefly describe the formalism used to evaluate inelastic
neutrino-nucleus cross sections at finite temperature. The results are
presented in section \ref{sect-Results}.  For each nucleus, the cross
sections are shown for several representative supernova
temperatures. Since the important quantity for supernova simulations
is the energy transfer from neutrinos to the core matter, we present
cross sections as a function of initial and final neutrino energies
for representative nuclei.  We finish the paper with a summary and
conclusions in section \ref{sect-Summary}. There we also make some
remarks on the impact of our results on the matter heating rate in
supernova simulations.

\section{Models}
\label{sect-Models}

In neutrino-induced reactions the nucleus is excited by multipole
operators ${\cal O}_\lambda$ which scale like $(qR/\hbar c)^\lambda$, where
$R$ is the nuclear radius ($R \sim 1.2 A^{1/3}$ fm). As the momentum
transfer $q$ is of the order of the neutrino energy $E_\nu$,
neutrino-nucleus reactions involve multipole operators with successively
higher rank $\lambda$ as the neutrino energy increases \cite{Kolbe03}.

Inelastic scattering of low-energy neutrinos
off nuclei is dominated by allowed ($\lambda=0$)
Gamow-Teller transitions.
The GT response is in turn sensitive to nuclear structure effects and
hence its contribution to the cross section has to be derived from a
model which is capable to describe both the relevant nuclear structure and
the correlation effects. This model is the diagonalization shell model
\cite{Caurier03}. The finite temperature in a supernova
environment implies that the scattering occurs on a thermal nuclear
ensemble rather than the ground state. However, the relevant supernova
temperatures ($T \le 2 $ MeV) ensure that mainly states at modest
nuclear excitation energies, which are also reasonably well described
within large-scale shell model approaches,  are present in
the thermal ensemble. As we will show below, finite-temperature effects
are only relevant for low neutrino energies $(E_\nu \simleq 15$ MeV).
Thus we expect that GT transitions will also dominate the 
neutrino scattering cross
section on the thermally excited nuclear states at these neutrino
energies.

Higher multipoles contribute to the cross section for
larger neutrino energies. For each of these multipoles
the response of the operator will be fragmented over many nuclear
states. However, most of the strength resides in a collective
excitation, the giant resonance, whose centroid energy grows with
increasing rank $\lambda$
roughly like $\lambda \hbar \omega \approx
41 \lambda / A^{1/3}$ MeV. As the phase space prefers larger final
neutrino energies, the average nuclear excitation energy grows
noticeably slower than the initial neutrino energy. As a consequence,
initial neutrino energies are noticeably larger than the energy of a
giant resonance,
when the latter will contribute to the neutrino-nucleus cross section.
Fortunately, the neutrino-nucleus cross section depends then mainly on
the total strength of the multipole excitation and its centroid energy,
and not on the detailed energy distribution of the strength (as it is
for the Gamow-Teller response at low neutrino energies).
We will derive the higher multipole contributions to the neutrino-nucleus
cross section within the RPA, making use of the fact
that the RPA describes the energy centroid and the total 
strength of multipoles other than the Gamow-Teller quite well.

Kolbe {\em et al\/} \cite{HybridModel} proposed this hybrid model,
in which contributions of the allowed transitions $(\lambda=0$)
to the neutrino-nucleus cross section are derived within the shell
model, and the other multipole responses ($\lambda >0$) are calculated 
within the RPA.
It has been shown that the model reproduces the experimental
$(\nu_e,e^-)$ cross section on $^{56}$Fe, induced by neutrinos with a
Michel energy spectrum \cite{HybridModel}.  Toivanen {\em et al\/}
applied this model to various neutrino-induced
reactions on iron isotopes at
the temperature $T=0$ \cite{Toivanen01}.  
Unfortunately, no inelastic
neutrino-nucleus scattering data for nuclei exists, except for the 
transition
from the ground state to the isospin $T=1$ state at $E_x=15.11$ MeV in
$^{12}$C \cite{Karmen}. However, it was recently shown
that high-precision 
data on the magnetic dipole strength distribution, obtained
from inelastic electron scattering on
spherical nuclei, like \up50Ti, \up52Cr and \up54Fe, give 
the required information
about the Gamow-Teller strength distribution
for these nuclei and thus strongly
constrain inelastic low-energy neutrino-nucleus cross sections
\cite{Langanke04}.
The diagonalization shell model reproduces the
electron scattering data very well, validating this model for
calculation of low-energy inelastic neutrino-nucleus cross
sections. Alternatively, the relevant $GT_0$ strength could be
obtained through a complete isospin decomposition of the $GT_-$
strength measured in $(p,n)$-type reactions,
see~\cite{Fujita96}. Nevertheless a direct measurement
of neutrino-nucleus cross sections for selected nuclei relevant for
supernova simulations is desirable. Such measurements could be
performed with a dedicated detector at a neutron spallation source 
\cite{SNS}.

In the hybrid model, the total reaction cross section 
consists then of two parts 
\begin{equation}
  \sigma_\nu^{tot}(E_\nu)\, =\, 
  \sigma_\nu^{sm}(E_\nu)\, 
  +
  \sigma_\nu^{rpa}(E_\nu),
\label{eq-sigma-tot}
\end{equation}
where the shell model part, 
$\sigma_\nu^{sm}$, 
describes the Gamow-Teller contributions to
the cross section, while the contributions from all other multipoles are
comprised in the RPA part,
$\sigma_\nu^{rpa}$.
The total cross section (denoted as `cross section' in the following) is
a function of the initial neutrino energy. We also calculate
cross sections as a function of initial and final neutrino
energies (called `differential cross sections' below), obtained by gating 
$\sigma_\nu^{tot}(E_\nu)$
on a range of outgoing neutrino energies
$[E_\nu'-\Delta E_\nu, E_\nu')$
for a given energy of the incoming neutrino, $E_\nu$.
In principle, supernova simulations require the inelastic neutrino cross
sections also as a function of angle of the final neutrino.
The hybrid model allows the calculation of such
differential cross sections. However, we have omitted this degree of
freedom here, 
since supernova simulations
can already judge the importance
of inelastic
neutrino-nucleus scattering
on the basis of the differential cross sections and
assuming an isotropic angular distribution.

As discussed above, finite-temperature effects will modify the
Gamow-Teller contribution to the cross section.
However, an explicit calculation of the cross section 
by shell model diagonalization at finite
temperature ($T\simgeq1$ MeV) includes too many states to derive the
$GT$ strength distribution for each individual state and is hence
unfeasible. We, therefore, use the same strategy as in \cite{Sampaio02}. 
We split the
shell model cross section into parts describing i) neutrino
down-scattering ($E_{\nu}'\leq E_\nu$) and ii) up-scattering ($E_{\nu}'\geq
E_\nu$); here $E_{\nu}',$ $E_\nu$ are neutrino energies in the final and
initial states, respectively. For the down-scattering part we apply
Brink's hypothesis which states that for a given excited nuclear level $i$
the $GT$ distribution built on this state, $S_i(E)$, is the same as
for the ground state, $S_0(E)$, but shifted by the excitation energy
$E_i$: $S_i(E)=S_0(E-E_i)$. Brink's hypothesis was proved valid 
if many states contribute to the thermal nuclear ensemble
\cite{Langanke00}. Upon applying Brink's hypothesis, the
down-scattering part becomes independent of temperature and can be
solely derived from the ground state $GT$ distribution.
With this approximation, the Gamow-Teller (shell model) contribution to
the cross section becomes:
\begin{equation}
 \sigma_\nu^{sm}(E_\nu)= \frac{G_F^2}{\pi} 
 \left[ 
   \sum\limits_{f} E_{\nu,0{f}}^{\prime \,2}\, B_{0{f}} (GT_0)\,\,
  +\,\,
   \sum\limits_{if} E_{\nu,{if}}^{\prime \,2}\, B_{{if}}
 (GT_0)\, \frac{G_{i}}G 
 \right],
\label{eq-sm-cross-section}
\end{equation}
where $G_F$ is the Fermi constant, 
and $E'_{\nu, {if}}$ is the energy of the scattered neutrino,
$E'_{\nu,{if}}=E_\nu+(E_i-E_f)$,
with $E_i,$ $E_f$ denoting the initial and final nuclear
energies.
The $GT$ reduced transition probability between the initial and final
nuclear states are given by
\begin{equation}
  B_{if}(GT_0)\,=\,
  \left(\frac{g_A}{g_V}\right)_{e\!f\!f}^2 
  GT_0
  \,=\,
  \left(\frac{g_A}{g_V}\right)_{e\!f\!f}^2 
  \frac{
    | \langle i|| 
     \sum_k {\vec {\bm \sigma}}^k \bm{t}_0^k
      || f\rangle |^2
  }{(2J_i+1)}
\end{equation}
where the matrix element is reduced with respect to the spin operator
$\vec {\bm \sigma}$ only and the sum runs over all nucleons;
$\bm{t}_0$ is the zero-component of the isospin operator in a
spherical representation.
As required in $0\hbar\omega$ shell model calculations, 
the $GT$ matrix elements have to be scaled
by a constant quenching factor. We use \cite{Martinez96}
$$
  \left( \frac{g_A}{g_V} \right)_{e\!f\!f}\,
  =\, 0.74\,
  \left( \frac{g_A}{g_V} \right)_{bare}
$$
with $(g_A/g_V)_{bare}=-1.2599(25)$ \cite{gagv}.

The first term in eq.\ \refeq{eq-sm-cross-section} arises from Brink's
hypothesis.  By construction, this term does not allow
neutrino up-scattering. These contributions to the cross section are
comprised in the second term, where the sum runs over both initial ($i$)
and final states ($f$). The former have a thermal weight of
$G_i=(2J_i+1)\exp(-E_i/T)$, where $J_i$ is the angular momentum,
$E_i$ is the energy of the initial state, and $T$ is the temperature
in MeV;
$G=\sum_i G_i$ is the nuclear partition function.  The up-scattering
contributions are the more important i) the lower the nuclear
excitation energy (the Boltzmann weight in the thermal ensemble regulates 
the population of the states), ii) the
larger the $GT$ transition strength $B_{if}$, and iii) the larger the
final neutrino energy, $E'_{\nu,if}$. 
Guided by these general considerations we
approximate the second term in eq.\ \refeq{eq-sm-cross-section} by
explicitly considering GT transitions between nuclear states with
$E_i > E_f$, where the sum over the final nuclear states is restricted
to the lowest (four to nine) 
nuclear levels.  The respective $GT$ matrix
elements are determined by an 'inversion' of the shell model $GT$
distributions of these low-lying states, employing a Lanczos
diagonalization technique with 35-60 iterations per final angular
momentum and isospin.  Although the Lanczos states at moderate
energies $E \simgeq 4$ MeV represent the energy distribution of the strength,
rather than converged nuclear states, the chosen ensemble of final
nuclear states appears sufficient to approximate the nuclear partition
function at the temperatures studied here.  
The lowest shell model energies
were replaced by experimental data \cite{exp-spectrum}, whenever
available.

The contributions of higher multipoles with rank $\lambda >0$ to the
cross section, $\sigma_\nu^{rpa}$ in eq.\ \refeq{eq-sigma-tot}, were
calculated using a generalized version of the RPA, which allows for
partial occupancies of the single-particle orbits in the parent ground
state. 
We have taken the proton and neutron occupation numbers from
the Independent Particle Model, which has been shown to yield quite
similar cross sections for multipoles with $\lambda >0$ as an approach
where the shell model occupation numbers were used \cite{Toivanen01}.
The RPA energies were derived from an appropriate Woods-Saxon
potential, adjusted to reproduce the experimental values of the proton
and neutron separation energies.  As the residual interaction in the
RPA calculation we adopted a zero-range Migdal force. The RPA
formalism and its application to neutrino-nucleus reactions is
described in detail in Refs.\ \cite{Kolbe03,Kolbe92,Kolbe99}.

\section{Results}
\label{sect-Results}

Our study covers neutral-current neutrino reactions on the nuclei
$^{50-60}$Mn, 
$^{52-61}$Fe, 
$^{54-63}$Co, 
and 
$^{56-64}$Ni, which are quite abundant during 
the supernova evolution.
Furthermore, large-scale shell model calculations  
for the Gamow-Teller strength distributions can
be performed for these nuclei.
We present our results in three subsections.
The Gamow-Teller distributions for the ground states,
which determine the temperature-independent shell model contribution to the 
cross-section (first term in eq.\ \refeq{eq-sm-cross-section}) 
are discussed in subsection
\ref{ssect-GT0}. 
As we pointed out above, the thermal population of nuclear excited states
influences the cross sections.
We give examples of this dependence 
for three typical supernova temperatures in subsection
\ref{ssect-sigma-tot}. 
The differential cross sections
are discussed in subsection \ref{ssect-d2s}, where
we also compare the contributions of the down-scattering and
up-scattering neutrino processes.

The cross-sections were calculated for larger ranges
of temperature and neutrino energies than presented here. The full set
of data is available from the corresponding author upon request.

\subsection{$GT_0$ ground state transition strength}
\label{ssect-GT0}

\begin{figure}[tbp]
\centerline{\psfig{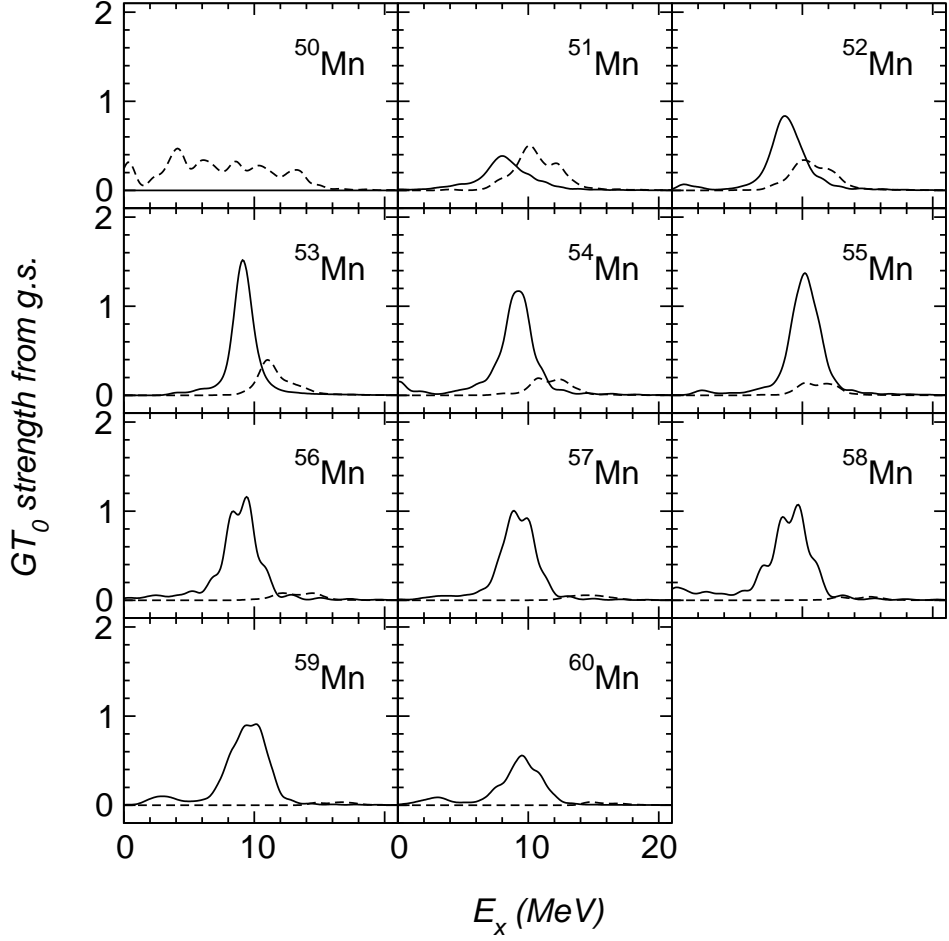}}
\caption{\label{fig-Mn-GT0} Gaussian-smoothed neutral Gamow-Teller
  ($GT_0$) 
distributions from the ground states in Mn isotopes. Solid lines indicate
$\Delta I=0$ strength, and dashed lines - $\Delta I=1$. 
$E_x$ is the nuclear excitation energy, $E_x=E_i-E_{gs}$.}
\end{figure}
\begin{figure}[tbp]
\centerline{\psfig{figure=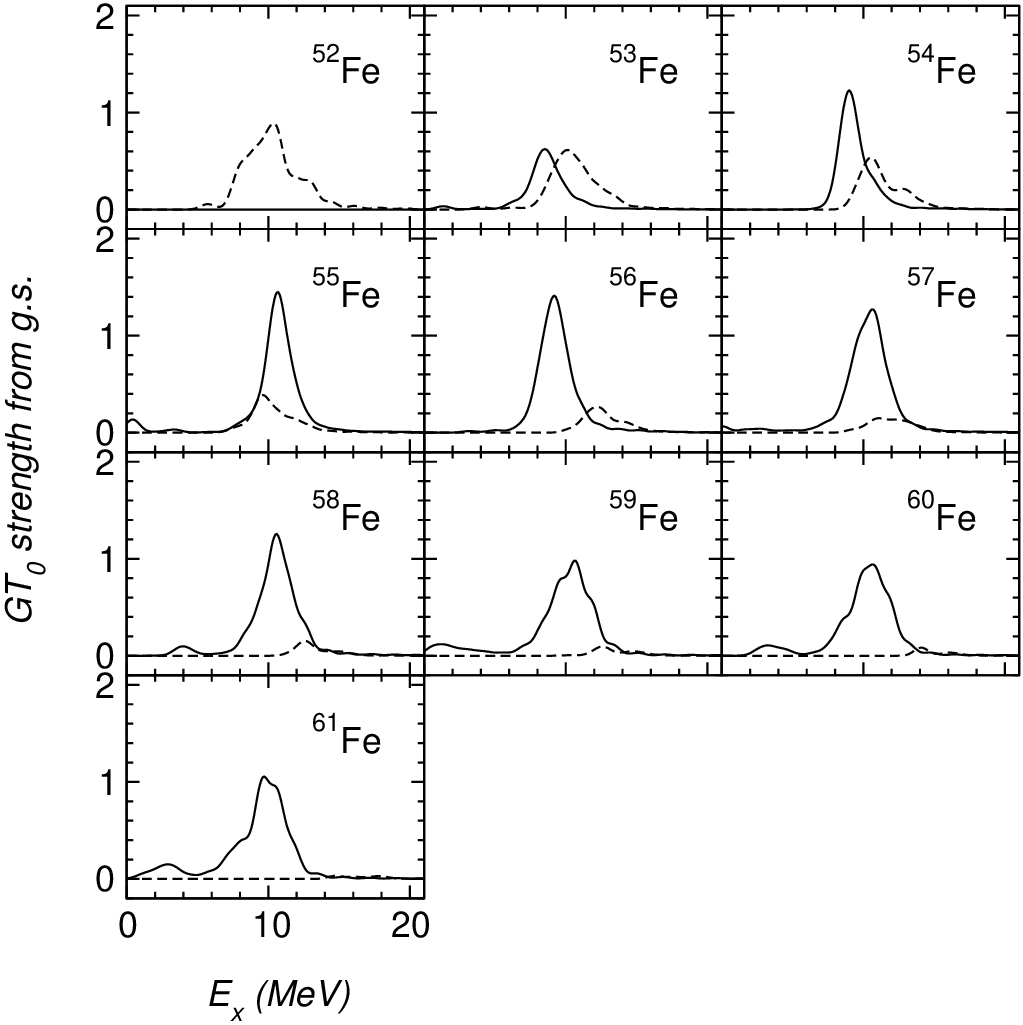,width=0.9\textwidth,angle=270}}
\caption{\label{fig-Fe-GT0} Same as Fig.\ \ref{fig-Mn-GT0}, but for Fe isotopes.}
\end{figure}
\begin{figure}[tbp]
\centerline{\psfig{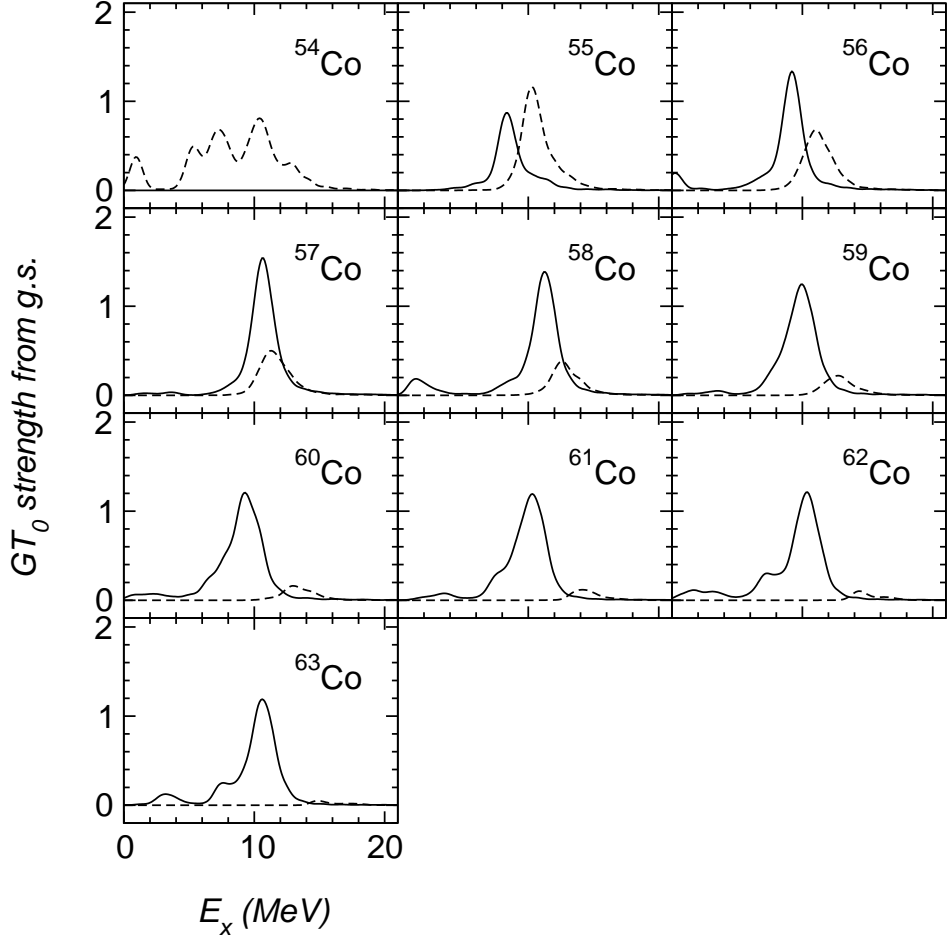}}
\caption{\label{fig-Co-GT0} Same as Fig.\ \ref{fig-Mn-GT0}, but for Co
  isotopes.}
\end{figure}
\begin{figure}[tbp]
\centerline{\psfig{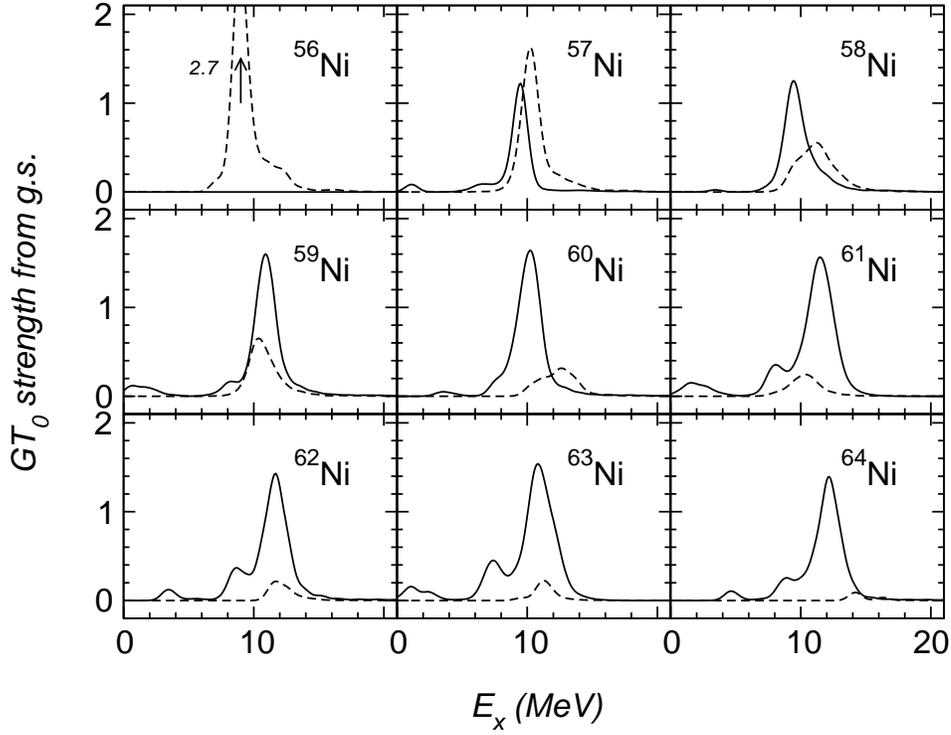}}
\caption{\label{fig-Ni-GT0} Same as Fig.\ \ref{fig-Mn-GT0}, but for Ni isotopes.
The distribution peak in
\up56Ni is at $E_x\sim9$ MeV and reaches the value of 2.7.}
\end{figure}
\begin{figure}[tbp]
\centerline{\psfig{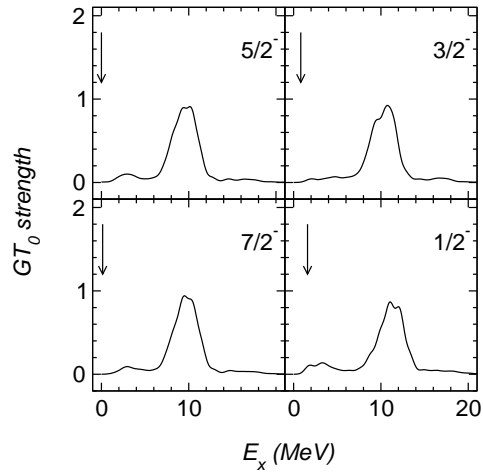}}
\caption{\label{fig-Mn59-GT0s} 
Gamow-Teller distributions from the 4 lowest states in \up59Mn. An arrow
indicates the calculated excitation energy of a state.
}
\end{figure}
\begin{figure}[tbp]
\centerline{\psfig{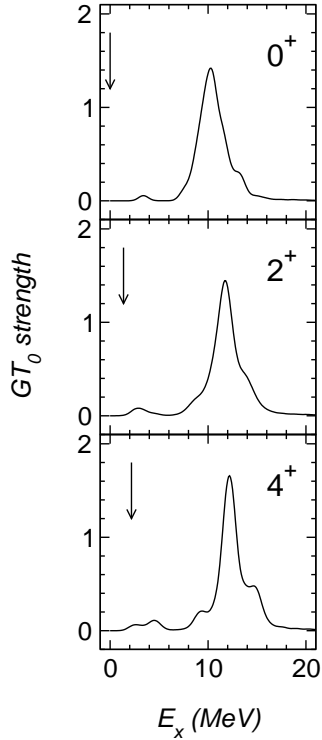}}
\caption{\label{fig-Ni60-GT0s} 
Gamow-Teller distributions for 3 states in \up60Ni. 
See also caption of Fig.\ \ref{fig-Mn59-GT0s}.
}
\end{figure}

The relevant nuclear structure information resides in the matrix elements
$B_{if}(GT_0)$,
that define the 
strength for the Gamow-Teller operator ${\vec {\bm \sigma}}\bm{t}_0$ 
between
initial and final states, see eq.\ \refeq{eq-sm-cross-section}. These
matrix elements were taken from 
large-scale shell model calculations in the
$pf$-shell using the computer code Antoine \cite{Antoine} and taking a
slightly modified version of the KB3 residual 
interaction \cite{Caurier99}.
Some of these calculations were reported in Ref.\
\cite{Caurier99} and were already used in several other
neutrino-nucleus reaction studies for both charged- and
neutral-current calculations: Toivanen {\it et al\/} \cite{Toivanen01}
presented the total cross sections on $^{52-60}$Fe for the temperature 
$T=0$ assuming several distributions of
supernova neutrinos and antineutrinos. Sampaio {\it et
al\/} \cite{Sampaio01} discussed neutrino and antineutrino 
absorption 
cross sections on 
$^{59,61}$Fe, $^{60,62}$Co, and $^{60,62}$Ni. 
The total cross-sections and normalized neutrino
spectra for neutral-current neutrino reactions on
$^{56,59}$Fe and $^{56,59}$Co were presented
in Ref.\ \cite{Sampaio02}.

In Figs.\ \ref{fig-Mn-GT0}-\ref{fig-Ni-GT0} 
we present the $GT_0$ strength distributions for the various isotope
chains; these strength distributions determine
the temperature-independent shell model contribution to the cross section
(the first term in eq.\ \refeq{eq-sm-cross-section}).  
The neutral
Gamow-Teller operator, 
${\vec {\bm \sigma}}\bm{t}_0$,
connects state $(J_i,I_i)$
with the states $J_f-J_i=0,\pm1$ (but not $J_i=J_f=0$) and
$I_f-I_i=0 (\Delta I=0),\pm1 (\Delta I=1)$. 
Thus $\Delta I=0$ transitions involve a change in
the angular momentum, $J_f-J_i=0,\pm1$, but not in isospin. 
The $\Delta I=1$ transition
may have a lower ($I_i-1$) or higher ($I_i+1$) isospin in addition to
a change in the angular momentum. 
As the odd-odd $N=Z$ nuclei $^{50}$Mn and $^{54}$Co have ground state
isospin $I=1$, both $\Delta I=1$ components can contribute to 
the GT distribution in these
nuclei. For all other nuclei,
$\Delta I=1$ allows only
$I_f=I_i+1$ transitions from the ground state. Furthermore,
there is no $\Delta I=0$ transition strength from the ground states
of all $N=Z$ nuclei, as the relevant
isospin Clebsch-Gordon coefficients 
($\langle 0010|00\rangle$ and
$\langle 1010|10\rangle$) vanish.
We have distinguished the two isospin components of the $GT_0$ strength
distribution in 
Figs.\ \ref{fig-Mn-GT0}-\ref{fig-Ni-GT0}, 
where solid lines refer to the $\Delta I=0$ part,
and dashed lines show the
transition strengths where the isospin changes by 1 unit.  

We observe from Figs.\ \ref{fig-Mn-GT0}-\ref{fig-Ni-GT0} that the peak
of the $\Delta I=0$ ($I_i\rightarrow I_i$) $GT_0$ strength is at 8-12
MeV, while the strongest $\Delta I=1$ ($I_i\rightarrow (I_i+1)$)
transitions lie a bit higher, in the energy range of 10-15 MeV. The
position of the centroid does not indicate a pronounced sensitivity to
the pairing structure of the ground state (see also
\cite{Langanke00,Toivanen01}).  The $GT_0$ strength for even-even
nuclei is mainly concentrated in the resonance at around $\sim 10$
MeV, while some low-lying strength in the Fe and Ni isotopes develops
once nucleons start to occupy higher orbitals. Odd-$A$ and
odd-odd nuclei usually show some low-lying strength.  The $GT_0$
strength distributions for the two odd-odd $N=Z$ nuclei $^{50}$Mn and
$^{54}$Co are significantly more fragmented than the distributions for
all other nuclei which is caused by the unusual isospin structure of
these two nuclei.

Differences of our $GT_0$ strength distributions
to those presented in \cite{Sampaio02,Toivanen01} are related
to different binning and smoothing. Furthermore, our figures
do not show the `elastic'
(i.e. at $E_x=0$) contribution to the strengths.

Our calculations of the down-scattering contribution to the
cross-section 
(the first term in eq.\ \refeq{eq-sm-cross-section}) 
is built on the Brink hypothesis. To test this assumption we plot in
Figs.\ \ref{fig-Mn59-GT0s} and \ref{fig-Ni60-GT0s} 
the $GT_0$ strength distribution for a few
low-lying
states, adopting the nuclei \up59Mn and
\up60Ni as examples. 
The figures support the validity of the Brink hypothesis for the
centroid of the strength distribution, in agreement with the original
formulation of the hypothesis for the collective states. We note,
however, that the Brink hypothesis is not applicable for low-lying
transitions of single-particle nature, as has already been discussed
in Ref.\ \cite{Langanke00}.

\subsection{Thermal total cross-sections}
\label{ssect-sigma-tot}

\begin{figure}[tbp]
\centerline{\psfig{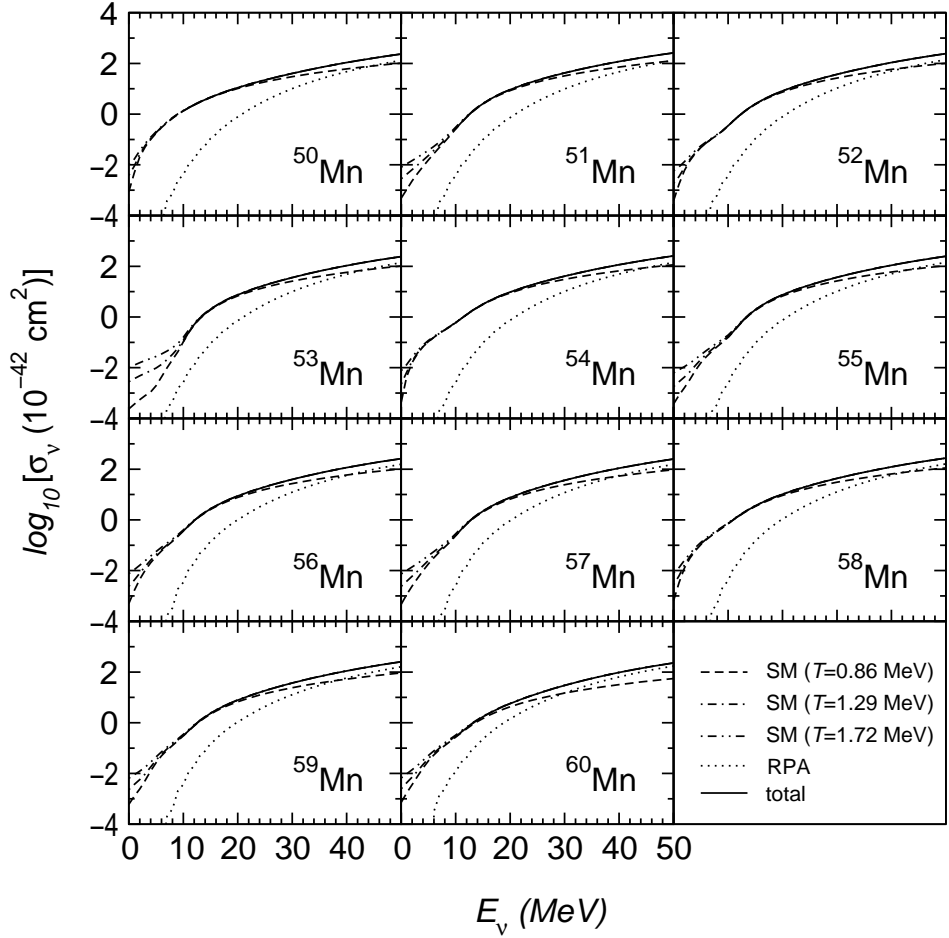}}
\caption{\label{fig-Mn-xsect} Neutral current neutrino-nucleus inelastic
cross-sections for Mn isotopes at different temperatures. Also shown 
is the RPA contribution to the cross section.
At low neutrino energies the total cross section
coincides with the shell model contributions.}
\end{figure}

\begin{figure}[tbp]
\centerline{\psfig{figure=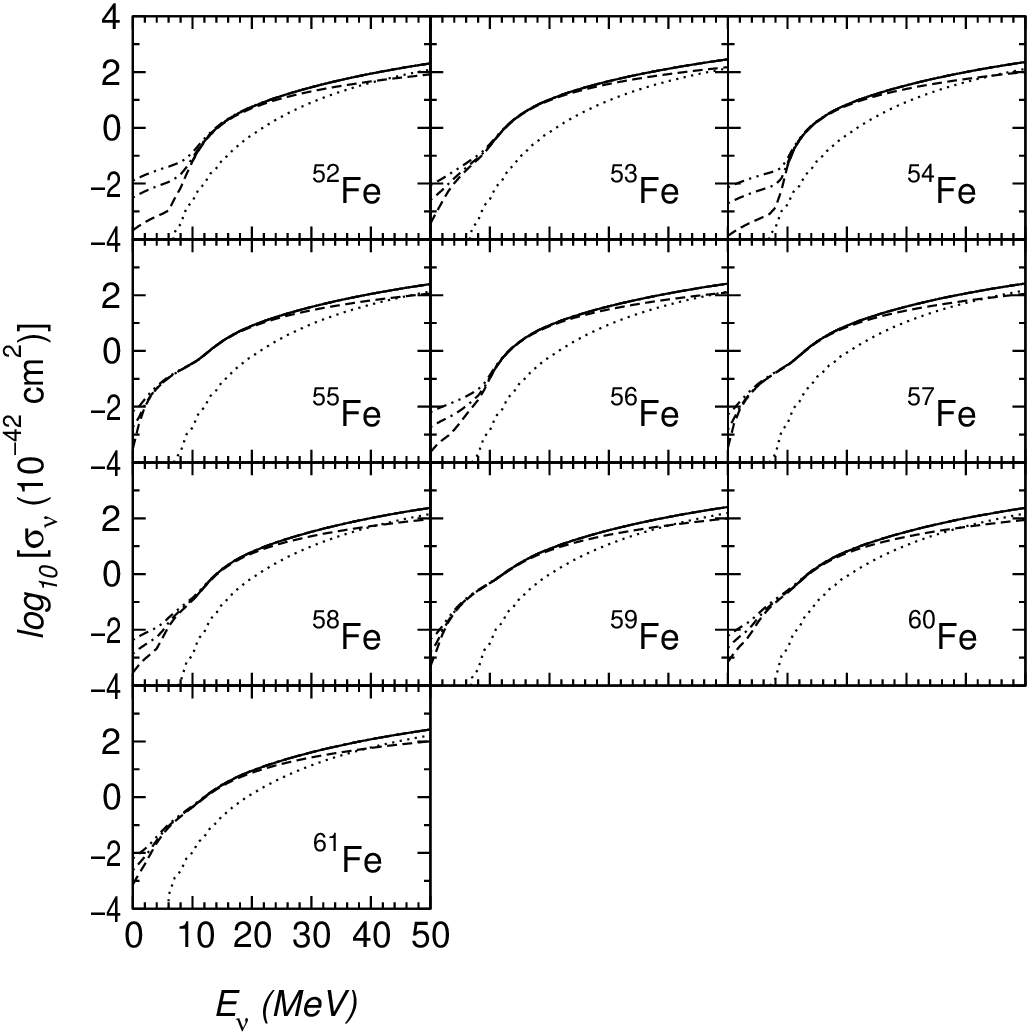,width=0.9\textwidth,angle=270}}
\caption{\label{fig-Fe-xsect} Same as Fig.\ \ref{fig-Mn-xsect}, but for Fe isotopes.}
\end{figure}
\begin{figure}[tbp]
\centerline{\psfig{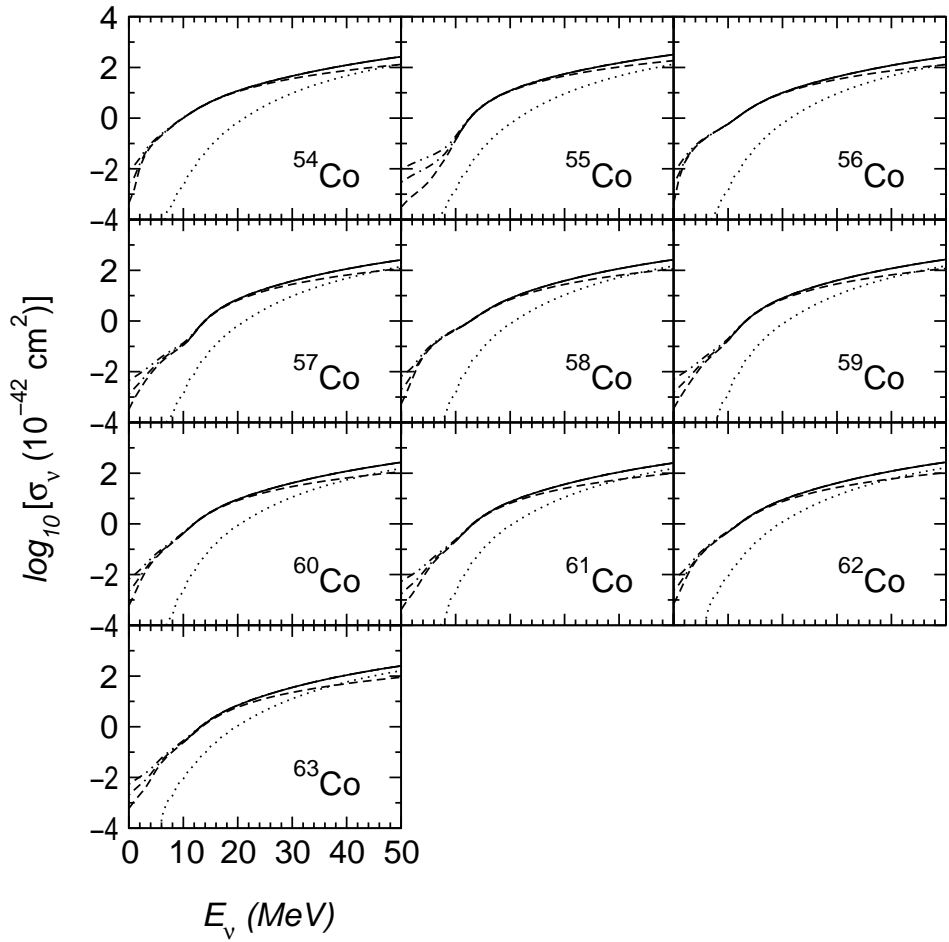}}
\caption{\label{fig-Co-xsect} Same as Fig.\ \ref{fig-Mn-xsect}, but for Co isotopes.}
\end{figure}
\begin{figure}[tbp]
\centerline{\psfig{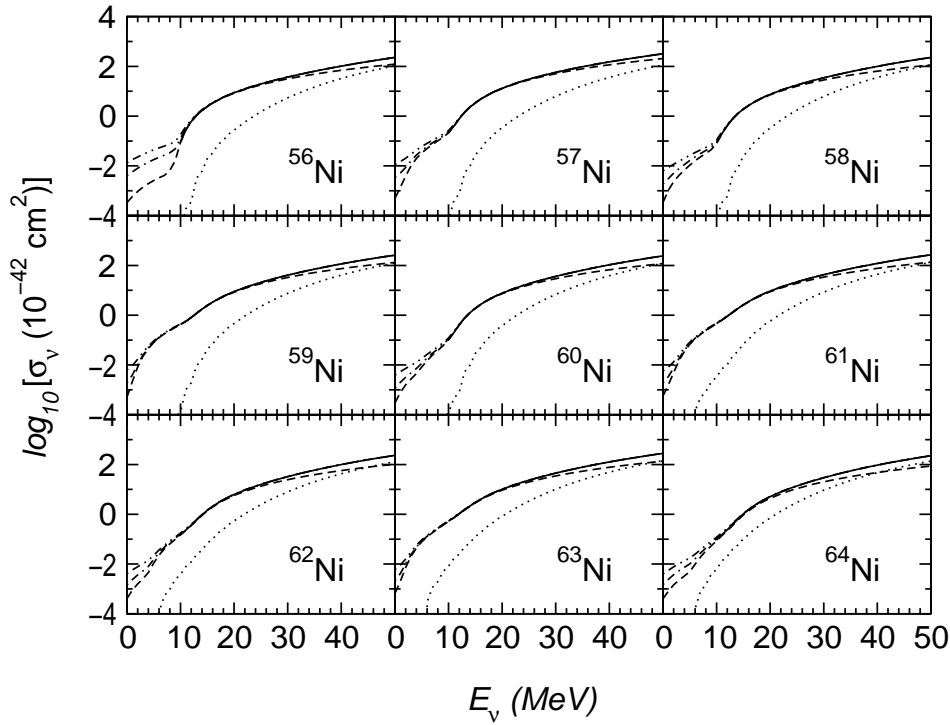}}
\caption{\label{fig-Ni-xsect} Same as Fig.\ \ref{fig-Mn-xsect}, but
  for Ni isotopes.
}
\end{figure}

As outlined in section \ref{sect-Models}, 
the contributions to the cross-sections arising from
allowed transitions, $\lambda=0$, and higher multipoles,
$\lambda>0$, are calculated within two different models: the shell model and
RPA, respectively. 
Further, we assume that the contributions from the higher multipoles
are temperature-independent. Thermal effects, which are included
via Gamow-Teller deexcitations to the ground state and to low-lying
excited states, are included
in the shell model part of the cross section (second term in eq.\
\refeq{eq-sm-cross-section}). 
Figs. \ref{fig-Mn-xsect}-\ref{fig-Ni-xsect}
show the inelastic neutrino-nucleus cross sections
for three different nuclear temperatures
relevant for supernova physics:
$T=0.86$ MeV ($10^{10}$
K), 1.29 MeV ($1.5\times10^{10}$ K), and 1.72 MeV ($2\times10^{10}$
K). The temperature $T=0.86$ MeV corresponds to the condition 
in the core of a
presupernova model for a 15$M_\odot$ star ($\rho\sim10^{10}$ g/cm$^3$
\cite{Heger01}). The two other temperatures relate approximately to
neutrino trapping ($T=1.29$ MeV) and thermalization ($T=1.72$ MeV).
For use in supernova simulations we have prepared cross section files
for wider ranges of nuclear temperatures (from $T=0.4$ to 2.0 MeV)
and neutrino energies (from $E_\nu=0$ to 100 MeV). 

A previous study of neutral current neutrino-nucleus reaction
cross-sections \cite{Sampaio02} suggested that finite-temperature
effects are
only relevant at low neutrino energies,
$E_\nu\simleq 10$ MeV.
This is related
to the energy of the $GT_0$ centroid which is located
around 10 MeV. Once neutrinos have sufficienly large energies
to excite the $GT_0$ centroid, the cross section is dominated by this
transition. Furthermore, as for excited states the relative 
excitation energy of the
centroid ($E_f-E_i$) is about the same as for the ground state, the
cross section becomes independent of temperature once transitions to the
$GT_0$ centroid dominate the cross section. At high neutrino energies,
other multipoles contribute to the cross section as well, where
the excitation is again mainly due to the collective excitations.
For these collective excitations, the Brink hypothesis applies again and
the contributions of the forbidden multipoles to the cross section
becomes temperature-independent.

At low neutrino energies, however, the cross section is temperature
dependent and factors like
the density of low-lying states and the $GT_0$ transition strength
between such states become important.
In general, odd-$A$ and odd-odd nuclei have more low-lying states 
than even-even nuclei, where a gap between the ground state and
the first excited state exists due to the isovector pairing. 
Furthermore, the ground states of even-even nuclei have angular momentum
$J=0$. $GT_0$ transitions can only connect these ground states to
$J=1$ states which usually exist at moderate excitation energies.
This angular momentum mismatch
creates a gap in the $GT_0$ strength distribution, which
translates into an energy threshold for inelastic neutrino scattering
at zero temperature. $GT_0$ transitions for excited states usually do
not show such a gap, as due to the increased density of states in their
vicinity the angular momentum mismatch is also absent.
As a consequence, there is no neutrino threshold energy at finite
supernova temperatures and low-energy inelastic neutrino cross sections on
even-even nuclei are quite sensitive to temperature.
As $E'_\nu < E_\nu$ in inelastic scattering on a
ground state, a neutrino threshold energy also exists in
odd-$A$ and odd-odd nuclei. However, as these nuclei
have a higher density of low-lying states and an 
angular momentum mismatch is usually missing, 
the energy threshold for odd-$A$ and odd-odd nuclei
is much smaller than for even-even nuclei. As a consequence, 
generally one expects
that temperature effects play a less important role for odd-$A$ and
odd-odd nuclei than for even-even nuclei \cite{Sampaio02}. 
However, due to nuclear structure effects there are exceptions to
this general rule.
The nuclei
\up53Mn (see Fig.\ \ref{fig-Mn-xsect}) and \up55Co (Fig.\
\ref{fig-Co-xsect}) are good examples of odd-even nuclei, 
where thermal effects are as important as in
the even-even nucleus
\up56Fe (Fig.\ \ref{fig-Fe-xsect}). 
This is related to the fact that 
\up53Mn and \up55Co have a closed $f_{7/2}$
neutron shell in the non-interacting shell model picture, which
reduces the 
density of low-lying states significantly.

We note that thermal effects can increase the low neutrino energy
cross sections for even-even nuclei (and the closed neutron shell
nuclei $^{53}$Mn and $^{55}$Co) by up to two orders of magnitude as
the temperature raises from 0.86 MeV to 1.72 MeV.  However, this
effect is noticeably milder in nuclei like \up60Fe, where the $GT_0$
strength distribution for the ground state exhibits some low-lying
strength.  We discuss the relative importance of thermal effects in a
greater detail in the next subsection.

\begin{figure}[tbp]
\centerline{\psfig{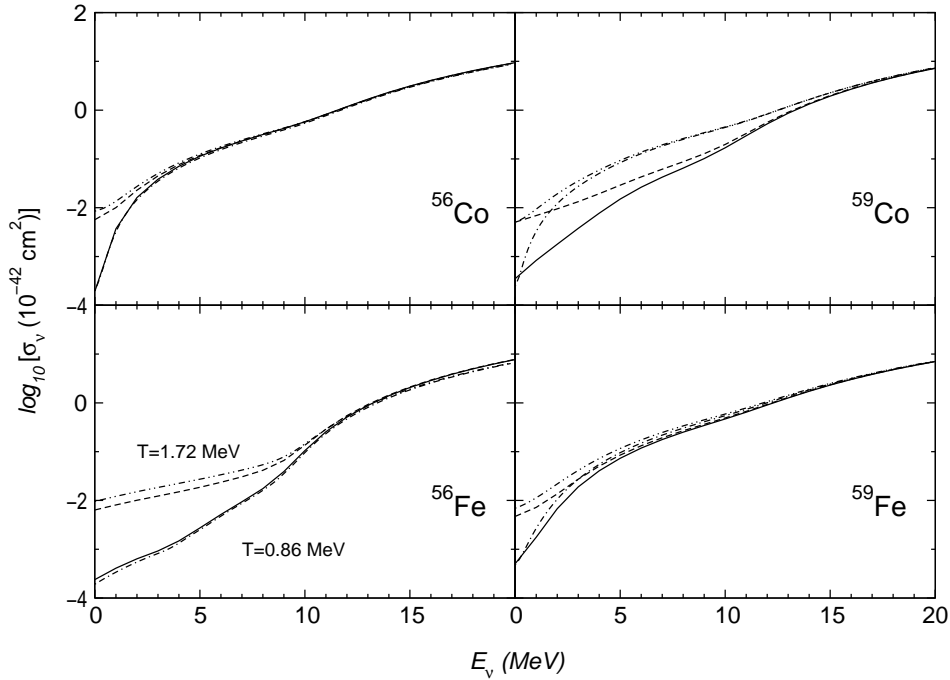}}
\caption{\label{fig-Sampaio-comparison} Comparison of the calculated
  shell model cross sections with those presented in Fig.\ 2 of Ref.\
  \cite{Sampaio02}. The solid and dashed
  lines show the current calculations at temperatures $T=0.86$ and 1.72
  MeV, respectively; the dash-dotted and dash-double-dotted lines show
  results from Ref.\ \cite{Sampaio02} for the same temperatures.
  }
\end{figure}

Sampaio {\it et al\/} studied inelastic neutrino scattering
off $^{56,59}$Fe and $^{56,59}$Co
\cite{Sampaio02}.
In Fig.\ \ref{fig-Sampaio-comparison}
we compare our results with those presented in 
Ref.\ \cite{Sampaio02}.
There are three differences
between our study and the previous one:
we do not include elastic transitions,
replaced our calculated low-excitation energies by experimental values
whenever available
\cite{exp-spectrum}, and calculated the value of the total partition
function $G$ more accurately.
As \cite{Sampaio02} only considered Gamow-Teller contributions to the
cross sections, 
we omit the contributions of higher multipoles to our cross sections 
in Fig.\ \ref{fig-Sampaio-comparison}.
The greatest difference is in the \up59Co cross-section,
attributed to
the omission of the
strong
elastic transition from the ground state.
(This omission is also the reason for the difference between
normalized neutrino spectra in \up59Co as shown in Figs.\ 3 and 4 of
Ref.\ \cite{Sampaio02} and our Fig.\ \ref{fig-Co59-EnuEnu} below.)

At low neutrino energies the cross-sections are
almost completely given by the allowed ($GT_0$) contribution. However,
contributions arising from forbidden multipoles become increasingly
important at larger neutrino energies.
We find that, for
$E_\nu=20$ MeV neutrinos, 
up to 18\% of the cross sections is due to the forbidden transitions
(in \up60Mn this is
24\%). 
For $E_\nu \sim 30$-50 MeV allowed and forbidden transitions contribute
about equally to the cross section, while at $E_\nu =100$ MeV the cross
sections are dominated by forbidden multipoles, with the $GT_0$
contributing about 15-20\%.

\subsection{Differential cross sections}
\label{ssect-d2s}

\begin{figure}[tbp]
\centerline{\psfig{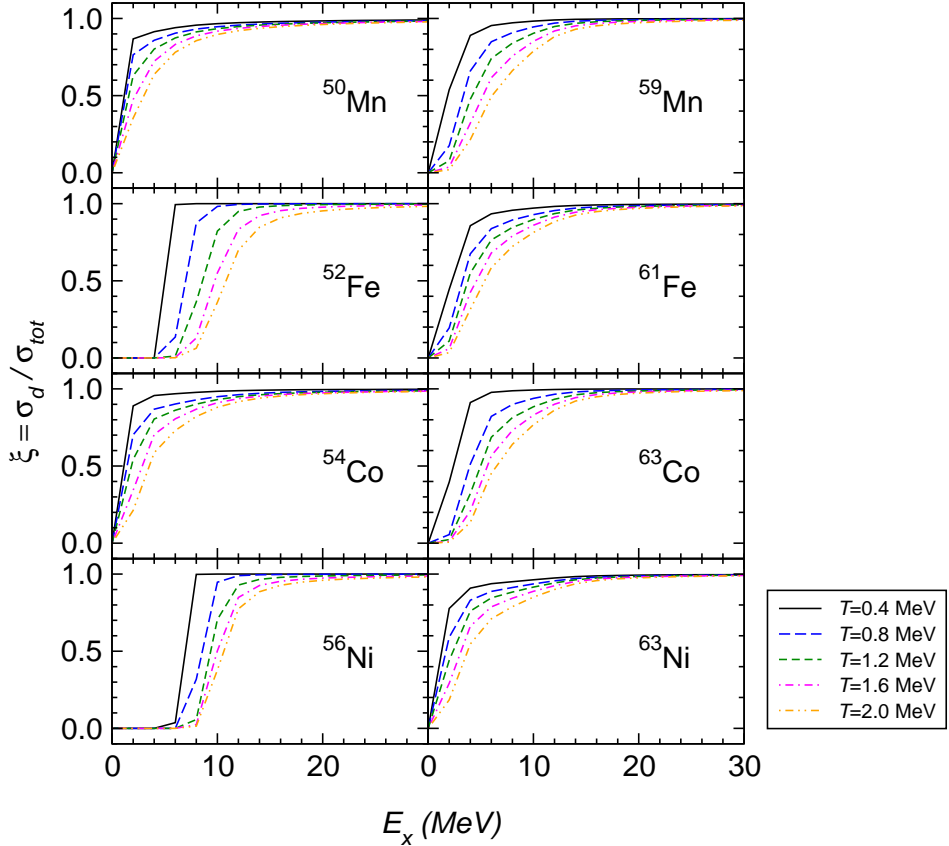}}
\caption{\label{fig-Ratio} 
  (Color online)
  Temperature dependence of the fraction of down-scattered
  neutrinos for selected nuclei: $^{50,59}$Mn, $^{52,61}$Fe,
  $^{54,63}$Co, and $^{56,63}$Ni. The sharp transitions are due to
  a coarse energy grid.}
\end{figure}

\begin{figure}[tbp]
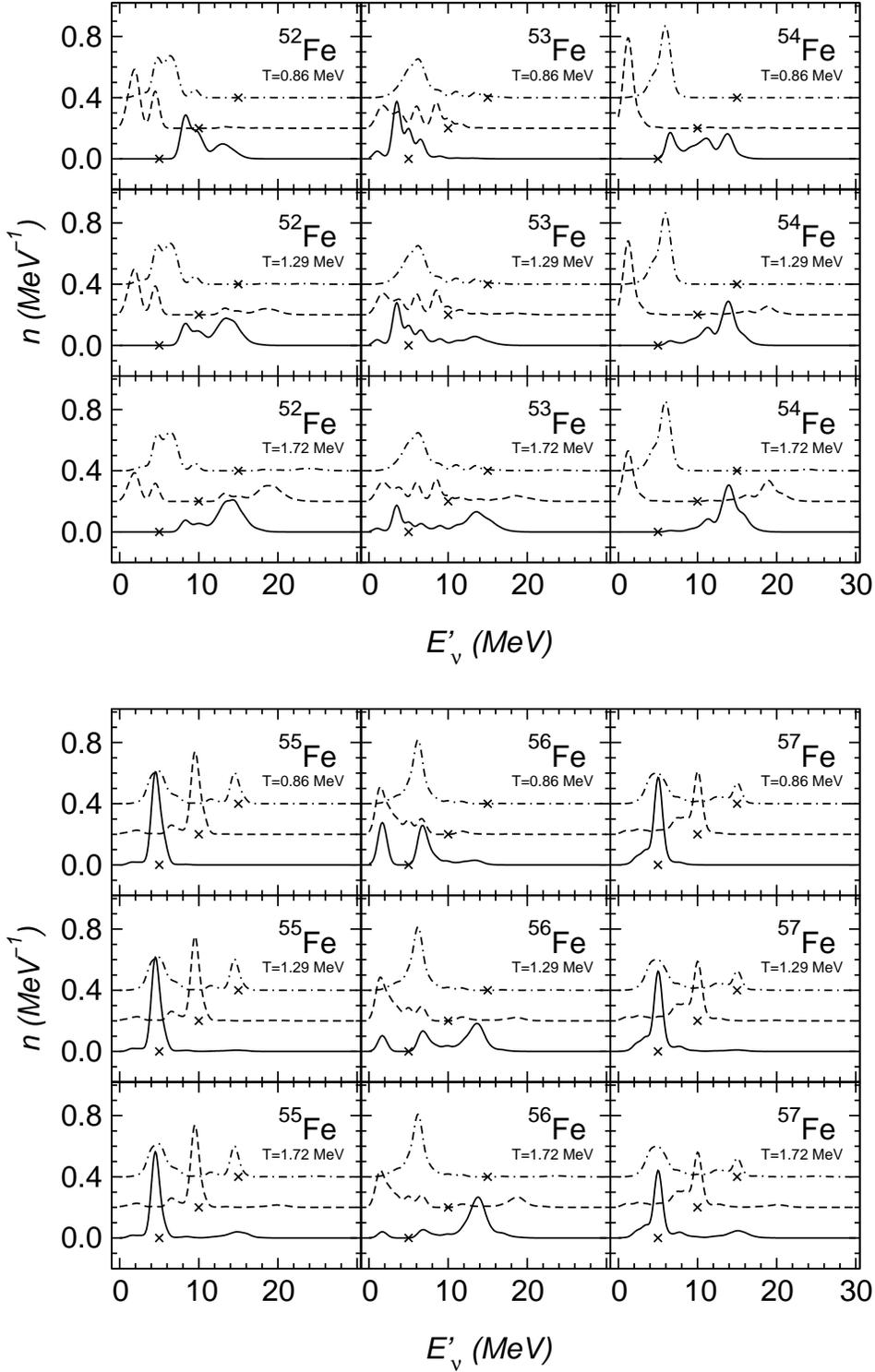

\centerline{\psfig{figure=figFe_nsp_52_54.eps,width=0.9\textwidth,angle=270}}

\vspace{0.5cm}
\centerline{\psfig{figure=figFe_nsp_55_57.eps,width=0.9\textwidth,angle=270}}

\caption{\label{fig-Fe52-EnuEnu} 
  Normalized spectra
  for final neutrino energies in inelastic neutrino scattering on
   $^{52-57}$Fe for three
  temperatures ($T=0.86$, 1.29, and 1.72 MeV) and three initial 
   neutrino energies: 
   $E_\nu=5$ MeV (solid line), 10 MeV (dashed line,
  all values shifted by 0.2), and 15 MeV (dash-dotted line, all values
  shifted by 0.4). Crosses correspond to the energy of the incoming
  neutrino. 
  }
\end{figure}
\begin{figure}[tbp]
\centerline{\psfig{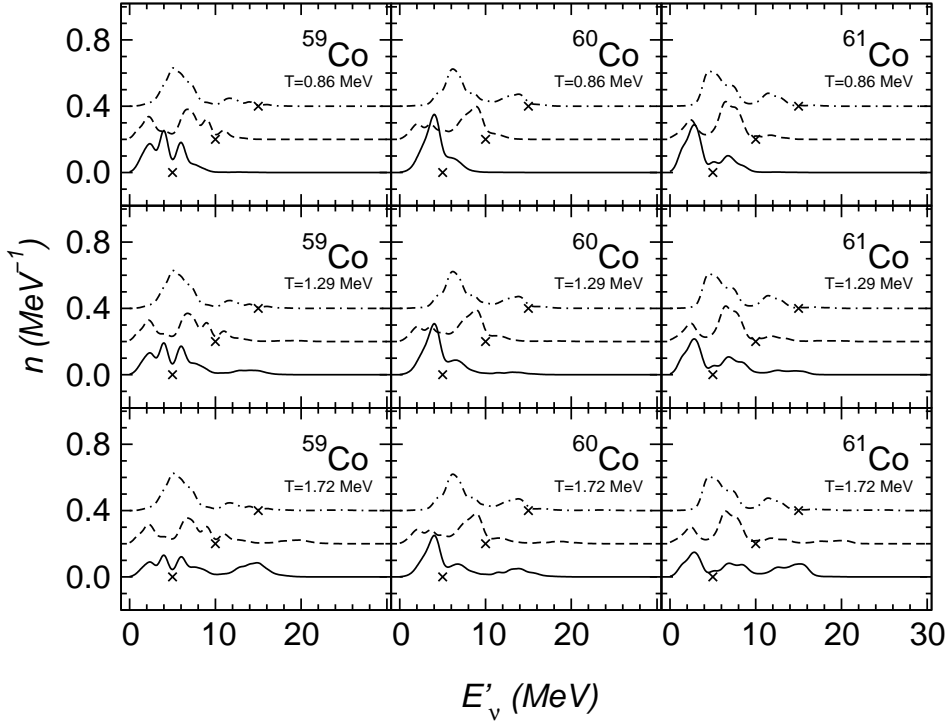}}
\caption{\label{fig-Co59-EnuEnu} Same as Fig.\ \ref{fig-Fe52-EnuEnu}
  but for $^{59,60,61}$Co}
\end{figure}
\begin{figure}[tbp]
\centerline{\psfig{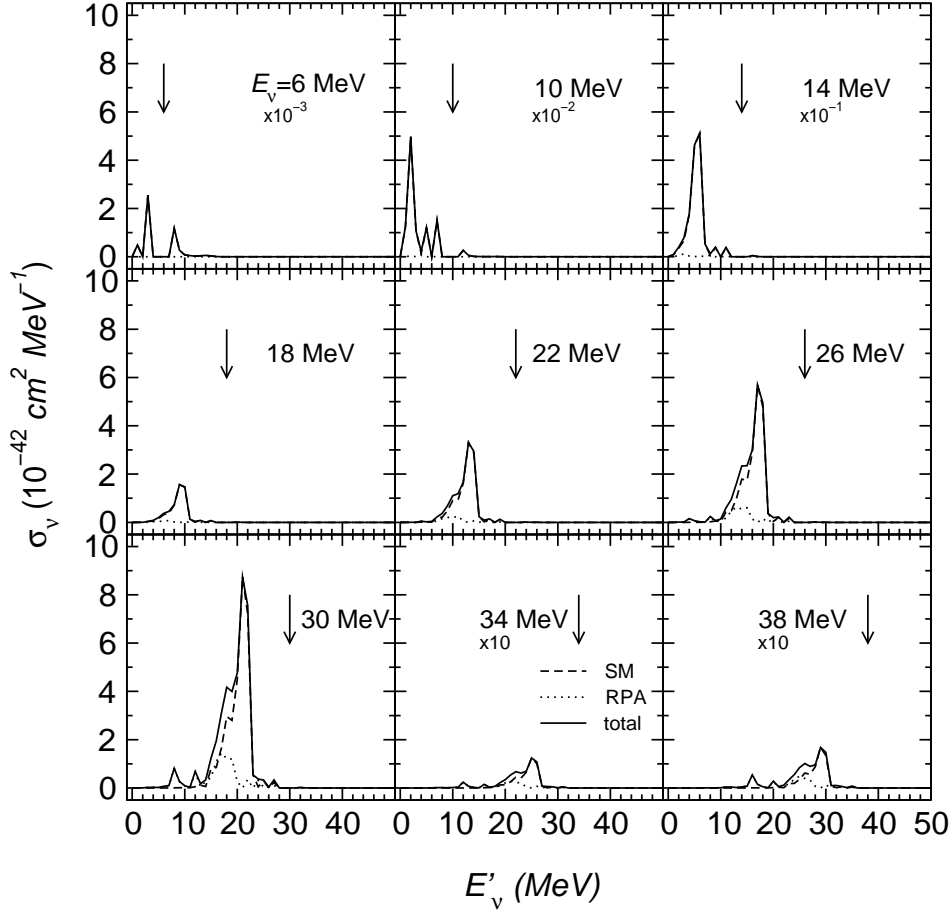}}
\caption{\label{fig-Fe56-partxsect}
  Differential
  cross-sections for inelastic neutrino scattering on \up56Fe for
  nine different initial neutrino energies and
  $T=0.8$ MeV. Each plot contains three lines: the dashed line shows
  the shell model, the
  dotted line - the RPA contributions, and the solid line shows the sum. 
  For some $E_\nu'$ values, the entire cross section comes either from
  the shell model or RPA contributions alone. 
  In all plots the energy of the incoming neutrino
  is indicated by an arrow. In some plots the cross sections have been scaled
  by the given factors (see text).
  The bin size is 1 MeV.
  }
\end{figure}

Among all reactions induced by neutrinos during the supernova collapse
the rate for elastic scattering of (electron) neutrinos on nuclei is
largest. As this process involves momentum transfer (but no energy
transfer), it randomizes the neutrino escape path from the collapsing
core and ultimately leads to the neutrino trapping and the formation
of the homologuous core (see e.g. \cite{BruennHaxton}). Thermalization
of neutrinos and supernova matter is made possible by neutrino
processes involving energy exchange.  As mentioned above, inelastic
neutrino-electron scattering is until now considered as the dominating
process, but inelastic neutrino-nucleus scattering, ignored so far,
could be quite important as well \cite{BruennHaxton}. We have
calculated the relevant partial differential cross sections for the
various Mn, Fe, Co, and Ni isotopes, covering initial neutrino
energies in the interval between $E_\nu=2$ and 60 MeV in 2 MeV steps and in
the interval $E_\nu=60$-100 MeV in 4 MeV steps. The cross sections are
then gated into 1 MeV bins for the final neutrino energies for
$E_{\nu}'<50$ MeV, and 2 MeV bins for $E_{\nu}'=50$-100 MeV. While a
table with a complete data set for all 40 individual nuclei is
available upon request from the corresponding author, we would like to
discuss general results and a few examples for the partial cross
sections in this subsection. For given neutrino energies we will
present our results as spectra for the neutrinos in the final
state. 

As discussed in the previous subsection,  inelastic neutrino-nucleus
cross sections for
low-energy neutrinos are increased by thermal
effects.
At higher neutrino energies
$E_\nu\simgeq10$ MeV, the main contribution to the cross-section comes
from the Gamow-Teller excitation, which resides at excitation energies
around 10 MeV.
Under these circumstances neutrino
down-scattering becomes the dominating process. Temperature effects
(neutrino up-scattering) are no longer important at $E_\nu\simgeq15$
MeV.
This general behavior can be visualized by introducing the ratio $\xi$
of the down-scattering cross section,
$\sigma_d$, and the total cross section, $\sigma_{\rm tot}$:
\begin{equation}
  \xi\,=\,\frac{\sigma_d}{\sigma_{\rm tot}}\, =\,
  1 - \frac{\sigma_u}{\sigma_{\rm tot}},
\end{equation}
where the up-scattering cross section
$\sigma_u(E_\nu)$ is given by the second term in eq.\
\refeq{eq-sm-cross-section}.

In Fig.\ \ref{fig-Ratio} 
we plot the ratio $\xi$ as a function of the initial neutrino energy
for selected temperatures and nuclei. As expected, 
the ratio $\xi\sim1$ is usually 
obtained for $E_\nu=15$ MeV
and temperature effects
become unimportant.
(For some nuclei, like \up52Fe and \up56Ni, temperature effects persist
to neutrino energies slightly larger than 15 MeV.)
At low neutrino energies we observe differences
between the individual nuclei, caused by the nuclear structure effects
discussed above.

In the previous subsection we outlined that even-even nuclei
(and certain others) have sizable energy thresholds for inelastic
neutrino scattering off the ground state. For these nuclei thermal
effects are more important and, obviously, they
are more likely to up-scatter low-energy neutrinos than those where
the threshold is virtually absent. 
This can be seen in Fig.\ \ref{fig-Ratio}, and we also
demonstrate this behavior for the nuclei
$^{52-57}$Fe and  $^{59-61}$Co in Figs.\
\ref{fig-Fe52-EnuEnu} and \ref{fig-Co59-EnuEnu}.  For each nucleus we
present the normalized-to-unity neutrino spectra
at three different
temperatures ($T=0.86$, 1.29, and 1.72 MeV) and, for each temperature, 
for three different initial neutrino energies are shown:
$E_\nu=5$, 10, and 15 MeV. 

As the spectra in Figs.\ \ref{fig-Fe52-EnuEnu} and
\ref{fig-Co59-EnuEnu} are normalized, the increase in absolute
magnitude with temperature is not reflected, but can be read off  Figs.\
\ref{fig-Mn-xsect}-\ref{fig-Ni-xsect}.
We remind here that in our formulation of the total cross section,
only the up-scattering processes
($E_\nu'>E_\nu$) are temperature-dependent.
The absolute value of the neutrino up-scattering 
cross section increases with temperature since a larger fraction
of nuclear excited
states with increasing phase space can be populated. 
Consequently, the relative contribution of up-scattered neutrinos in the
spectra increases with temperature and, for a given $E_\nu$, the spectra
of up-scattered neutrinos are
getting wider as temperature increases.

The missing low-energy strength in the ground state 
$GT_0$ distributions for the
even-even isotopes $^{52,54}$Fe (Fig.\ \ref{fig-Fe-GT0}) implies that
these nuclei have only 
a very few states which can be excited by a $E_\nu=5$
MeV neutrino. Therefore the neutrino distributions for these nuclei 
(Fig.\ \ref{fig-Fe52-EnuEnu}) are
dominated by up-scattered neutrinos even for temperatures as low as
$T=0.86$ MeV. 
The up-scattering process dominates also in \up56Fe at
$T=1.29$ and 1.72 MeV, but it is about equally significant as
down-scattering at $T=0.86$ MeV.
As expected from Fig.\ \ref{fig-Ratio},
a fraction of the $E_\nu=10$ MeV neutrinos is
up-scattered by even-even iron isotopes. For $E_\nu=15$ MeV neutrinos,
temperature effects are unimportant and nearly all neutrinos are
down-scattered.
As temperature effects are in general less important for odd-$A$ nuclei,
inelastic neutrino scattering on the odd  
isotopes $^{53,55,57}$Fe up-scatter a much smaller
fraction of 
neutrinos, even at 
$E_\nu=5$ MeV, than scattering on even iron isotopes.
Similarly, up-scattering is also relatively unimportant for odd-odd nuclei.
This is demonstrated in Fig.\
\ref{fig-Co59-EnuEnu}, where we show the normalized neutrino spectra
for the odd-$Z$ isotopes
$^{59-61}$Co. Although present for $E_\nu=5$ MeV neutrinos,
neutrino up-scattering is significantly less pronounced than for the
even-even iron isotopes.
When comparing the current neutrino spectrum 
for \up59Co with Fig.\ 3 in Ref.\ \cite{Sampaio02}, one can observe
that there is no big $E_{\nu}'\approx E_\nu$ contribution, which is
related to the omission of the elastic contribution to the cross section
here.

In Fig.\ \ref{fig-Fe56-partxsect} we show the (unnormalized) partial
cross sections for inelastic neutrino scattering off $^{56}$Fe,
choosing $T=0.8$ MeV as a representative supernova temperature. The
partial cross sections cover the range of initial neutrino energies
between $E_\nu=6$ MeV and 38 MeV in 4 MeV steps.  Note that for a
clearer presentation, some of the cross sections are scaled (i.e.\ the
actual values are obtained by multiplying the shown cross sections by
the indicated factors, which, for example, is $10^{-3}$ for $E_\nu=6$
MeV).  At all shown initial neutrino energies the cross sections are
dominated by the $GT_0$ contributions. As a consequence, the largest
partial cross sections are found for $E'_\nu \sim (E_\nu -10\ {\rm
MeV})$ for $E_\nu > 10$ MeV, corresponding to the centroid of the
$GT_0$ distribution. For higher neutrino energies we observe the
appearance of structures in the partial cross sections at final
neutrino energies smaller than corresponding to the $GT_0$
centroid. These peaks are related to excitations of the giant
resonances in forbidden multipole transitions, which, as expected,
become more prominent with increasing initial neutrino energies.  To
make this relation more visible, the figure shows the individual
contributions of the allowed and forbidden multipole transitions to
the partial cross sections.

\section{Summary and conclusions}
\label{sect-Summary}

We performed studies of neutral-current neutrino-nucleus reactions for
four isotope chains: Mn, Fe, Co, and Ni. These studies are relevant for
supernova simulations and hence have to take the finite temperature
of the supernova environment into account.
In our
approach, we determine the allowed contributions to the cross sections
from large-scale diagonalization shell model evaluations of the 
Gamow-Teller $GT_0$ response. Contributions to the cross sections,
arising from other multipoles, were calculated within
the random-phase approximation. We have calculated the $GT_0$
distributions for many nuclei for the first time. Our results indicate
a systematic trend in these distributions as they are usually dominated
by a collective excitation whose centroid is located at excitation
energies around 10 MeV in the nuclei studied here.
Finite temperature effects on the cross sections
are only relevant at rather small neutrino
energies and were approximated in our approach by considering
$GT_0$ transitions from thermally populated states to the ground
state and excited states at low nuclear excitation energies.

We
confirm the conclusions of Ref.\ \cite{Sampaio02} that finite temperature
effects enhances the inelastic scattering cross-sections for
low energy neutrinos. We also find that
temperature effects become negligible once the energy of the initial
neutrino is large enough to allow for transitions to the centroid of
the $GT_0$. Furthermore, due to nuclear structure effects the
temperature dependence is in general largest for even-even nuclei, and
among the nuclei studied here, also for the two odd-$A$ nuclei with closed
$f_{7/2}$ neutron shells ($^{53}$Mn and $^{55}$Co).

One goal of our study was to produce
inelastic neutrino-nucleus cross sections 
suitable for use in
supernova simulations. However, 
in the unshocked regions of supernova where
this process is relevant, many nuclei are present
in the  matter composition with sizable abundance. Supernova
simulations which contain detailed neutrino transport, 
however, represent this matter composition by  protons,
neutrons, $\alpha$ particles and `average nuclei' which simulate all
heavier nuclei, see e.g.\ \cite{Lattimer91}. 
For use in supernova simulations, 
the neutrino cross sections for the individual
nuclei have to be averaged over the supernova matter composition,
as it is, for example, done for stellar electron capture 
\cite{FFN82,rmp,Langanke03,Hix03}.
To simplify this procedure it has been proposed 
in Ref. \cite{Sampaio02} to determine the neutrino-nucleus cross sections
from an `average nucleus' which is chosen to approximate the 
matter composition.
Such a prescription is justified by our calculations
for larger neutrino energies, say at $E_\nu>15$ MeV, where the cross
sections for the various nuclei are rather similar.
However, the prescription appears to be less justified at low neutrino energies
where the cross sections for different nuclei show strong variations.
For these neutrino energies it is preferable to calculate the
total
neutrino-nucleus cross section 
as a 
compositionally
weighted average over the individual cross sections.
This procedure is facilitated as the supernova matter 
composition is given by Nuclear Statistical Equilibrium under most
conditions where the cross sections are needed. 

Inclusion of these rates in a supernova simulation is beyond the scope
of this paper. However,
to get an idea of
the importance of this process we estimated the heating rate in the
post-bounce region using results from a previous simulation \cite{Hix03}.  
Following Bruenn and Haxton \cite{BruennHaxton}, we
approximated the reaction cross sections on all nuclei in the
iron-shell by one nucleus. To check the variations in the produced
rates we took four different cases: \up55Co, \up56Co, \up55Fe, and
\up56Fe. The obtained heating rates were different by 40-50\%,
supporting the idea 
that the neutrino-nucleus heating rate in supernova is moderately
sensitive to the detailed composition of the region.

The astrophysical implications of these improved neutral-current
neutrino-nucleus reaction rates are currently under investigation.
Prior investigations by Bruenn and Haxton \cite{BruennHaxton} found
that the energy transfer
due to neutrino-nucleus scattering was comparable to (but
smaller than) that from neutrino-electron scattering during the infall
phase.  At later times in the simulation, Bruenn and Haxton found that
while neutrino-nucleus scattering dominated in the cooler iron-rich
regions, ${\bar \nu}_e$ capture on protons and neutrino scattering on
$^4$He were more important closer to the shock.  
Given the similarity of our cross sections to those used by Bruenn and
Haxton for the neutrino energy range of relevance for post-shock
heating
($E_\nu\simgeq15$ MeV), 
we expect their conclusions to stand. 
However, in addition to the
microscopic rates for the interactions themselves, these conclusions
also depend strongly on the composition which depends on the
hydrodynamic state and neutronization, since the matter in
question is in Nuclear Statistical Equilibrium.
Therefore these conclusions must be
revisited in light of recent improvements in supernova models.  Of
particular interest in this regard is the impact of revisions in the
treatment of nuclear electron capture \cite{Langanke03,Langanke00}
which have recently been shown to alter the thermodynamics and
neutronization throughout the collapsing stellar core
\cite{Hix03,Heger01}.
Improved cross-sections for $^{4}$He (see, however, \cite{Gazit04}), 
as well as
intermediate mass nuclei,
would be desirable, since these nuclei experience the intense neutrino
flux in the supernova as well.

\begin{ack}

We thank M.\ Liebend{\"o}rfer, A.\ Mezzacappa, S.W.\ Bruenn, and
O.E.B.\ Messer for useful discussions of supernova simulation results.
A.J.\ and W.R.H.\ are partially supported by the Department of Energy
through the Scientific Discovery through Advanced Computing (SciDAC)
program.  
W.R.H.\ is also supported in part by NSF under contract PHY-0244783.
K.L.\ and J.M.S.\ are partially supported by the Danish
Research Council.  G.M.P.\ is supported by the Spanish MCyT and by the
European Union ERDF under contracts AYA2002-04094-C03-02 and
AYA2003-06128.  J.M.S.\ acknowledges the financial support of the
Portugese Foundation for Science and Technology.  Oak Ridge National
Laboratory is managed by UT-Battelle, LLC, for the U.S.\ Department of
Energy under Contract No.\ DE-AC05-00OR22725 with UT-Battelle, LLC.

\end{ack}



\begin{thebibliography}{99}

\bibitem{BetheReview}
 H.A.\ Bethe, {\it Rev.\ Mod.\ Phys.\ \/} {\bf 62} (1990) 801.

\bibitem{Hix03JPG} 
 W.R.\ Hix, A.\ Mezzacappa, O.E.B.\ Messer, and S.W.\ Bruenn,
 \BibTitle{Supernova science at spallation neutron sources,}
{\it J.\ Phys.\ \/} {\bf G 29} (2003) 2523.


\bibitem{rmp}
  K.\ Langanke, and G.Mart{\'\i}nez-Pinedo,
  \BibTitle{Nuclear weak-interaction processes in stars,}
  {\it Rev.\ Mod.\ Phys.\ \/} {\bf 75} (2003) 819.

\bibitem{Kolbe03}
E.\ Kolbe, K.\ Langanke, G.\ Mart{\'\i}nez-Pinedo and P.\ Vogel,
  \BibTitle{Neutrino-nucleus reactions and nuclear structure,}
  {\it J.\ Phys.\ \/} {\bf G 29} (2003) 2569.

\bibitem{Mezzacappa01} 
  A.\ Mezzacappa, M.\ Liebend{\"o}rfer, O.E.B.\ Messer, W.R.\ Hix, 
    F.-K.\ Thielemann, and S.W.\ Bruenn,
 \BibTitle{Simulation of the Spherically Symmetric Stellar Core
 Collapse, Bounce, and Postbounce Evolution of a Star of 13 Solar
 Masses with Boltzmann Neutrino Transport, and Its Implications for
 the Supernova Mechanism,} 
 {\it Phys.\ Rev.\ Lett.\ \/} {\bf 86} (2001) 1935.

\bibitem{Rampp00}
  M.\ Rampp and H.-Th.\ Janka, 
  {\it Astrophys.\ J.\ \/} {\bf 539} (2000) L33.

\bibitem{Buras03}
  R.\ Buras, M.\ Rampp, H.-Th.\ Janka, and K.\ Kifonidis, 
  \BibTitle{Improved Models of Stellar Core Collapse and Still No
  Explosions: What Is Missing?,}
  {\it Phys.\ Rev.\ Lett.\ \/} {\bf 90} (2003) 241101.

\bibitem{Sampaio02}
 J.M.\ Sampaio, K.\ Langanke, G.\ Mart\'{\i}nez-Pinedo, and D.J.\
 Dean, 
 \BibTitle{Neutral-current reactions in the supernova environment,}
 {\it Phys.\ Lett.\ \/} {\bf B 529} (2002) 19.

\bibitem{Langanke03}
  K.\ Langanke, G.\ Mart{\'\i}nez-Pinedo, J.M.\ Sampaio, 
    D.J.\ Dean, W.R.\ Hix, O.E.B.\ Messer, A.\ Mezzacappa, 
    M.\ Liebend{\"o}rfer, H.-Th.\ Janka, and M.\ Rampp,
  \BibTitle{Electron Capture Rates on Nuclei and Implications for
  Stellar Core Collapse,}
 {\it Phys.\ Rev.\ Lett.\ \/} {\bf 90} (2003) 241102.

\bibitem{Hix03}
  W.R.\ Hix, O.E.B.\ Messer, A.\ Mezzacappa,
    M.\ Liebend{\"o}rfer, J.\ Sampaio, K.\ Langanke, 
    D.J.\ Dean, and G.\ Mart{\'\i}nez-Pinedo,
  \BibTitle{Consequences of Nuclear Electron Capture in Core Collapse
  Supernovae,}
 {\it Phys.\ Rev.\ Lett.\ \/} {\bf 91} (2003) 201102.

\bibitem{Bruenn85}
  S.W.\ Bruenn,
  \BibTitle{Stellar core collapse: Numerical model and infall epoch,}
  {\it Astrophys.\ J.\ Suppl.\ \/} {\bf 58} (1985) 771.

\bibitem{BruennHaxton}
 S.W.\ Bruenn and W.C.\ Haxton, 
 \BibTitle{Neutrino-nucleus interactions in core-collapse supernovae,}
 {\it Astrophys.\ J.\ \/} {\bf 376} (1991) 678.

\bibitem{Haxton88} W.C.\ Haxton, 
  \BibTitle{Neutrino heating in supernovae,}
 {\it Phys.\ Rev.\ Lett.\ \/} {\bf 60} (1988) 1999.

\bibitem{Caurier99}
  E.\ Caurier, K.\ Langanke, G.\ Mart{\'\i}nez-Pinedo, and F.\
  Nowacki,
  \BibTitle{Shell-model calculations of stellar weak interaction
  rates. I. Gamow-Teller distributions and spectra of nuclei in the
  mass range A = 45-65,}
 {\it Nucl.\ Phys.\ \/} {\bf A 653} (1999) 439.

\bibitem{Frekers}
  D.\ Frekers, 
  \BibTitle{Facets of (d,2He) charge-exchange reactions at
  intermediate energies,}
  {\it Nucl.\ Phys.\ \/} {\bf A 731} (2004) 76.

\bibitem{Hagemann}
  M.\ Hagemann\doetal{, A.M.\ van den Berg, D.\ De Frenne, V.M.\ Hannen,
  M.N.\ Harakeh, J.\ Heyse, M.A.\ de Huu, E.\ Jacobs, K.\ Langanke,
  G.\ Mart{\'\i}nez-Pinedo and H.J.\ W{\"o}rtche},
  \BibTitle{High-resolution determination of GT strength distributions
  relevant to the presupernova evolution using the (d,2He) reaction,}  
  {\it Phys.\ Lett.\ \/} {\bf B 579} (2004) 251.

\bibitem{Baeumer}
  C.\ B{\"a}umer\doetal{, A.M.\ van den Berg, B.\ Davids, D.\ Frekers, 
  D.\ De Frenne, E.-W.\ Grewe, P.\ Haefner, M.N.\ Harakeh, 
  F.\ Hofmann, M.\ Hunyadi, E.\ Jacobs, B.C.\ Junk, A.\ Korff, 
  K.\ Langanke, G.\ Mart{\'\i}nez-Pinedo, A.\ Negret, 
  P.\ von Neumann-Cosel, L.\ Popescu, S.\ Rakers, A.\ Richter, 
  and H.J.\ W{\"o}rtche},
  \BibTitle{High-resolution study of the Gamow-Teller strength
  distribution in 51Ti measured through 51V(d,2He)51Ti,}
  {\it Phys.\ Rev.\ \/} {\bf C 68} (2003) 031303R.

\bibitem{Kolbe92} 
  E.\ Kolbe, S.\ Krewald, K.\ Langanke, and F.-K.\
  Thielemann, 
  \BibTitle{Inelastic neutrino scattering on \up12C and \up16O
  above the particle emission threshold,}
  {\it Nucl.\ Phys.\ \/} {\bf A 540} (1992) 599.

\bibitem{Caurier03}
  E.\ Caurier, G.\ Mart{\'\i}nez-Pinedo, F.\ Nowacki, A.\ Poves
  and A.P.\ Zuker,
  \BibTitle{The Shell Model as Unified View of Nuclear Structure,}
  {\em nucl-th/0402046}.

\bibitem{HybridModel}
  E.\ Kolbe, K.\ Langanke, and G.\ Mart{\'\i}nez-Pinedo, 
  \BibTitle{Inclusive \up56Fe$(\nu_e,e^-)$\up56Co cross section,}
  {\it Phys.\ Rev.\ \/} {\bf C 60} (1999) 052801.

\bibitem{Toivanen01}
  J.\ Toivanen, E.\ Kolbe, K.\ Langanke, G.\ Mart{\'\i}nez-Pinedo, and
  P.\ Vogel,
  \BibTitle{Supernova neutrino induced reactions on iron isotopes,}
  {\it Nucl.\ Phys.\ \/} {\bf A 694} (2001) 395.

\bibitem{Karmen}
  B.\ Zeitnitz and KARMEN Collaboration,
  \BibTitle{KARMEN - NEUTRINO PHYSICS AT ISIS,}
  {\it Prog.\ Part.\ Nucl.\ Phys.\ \/} {\bf 32} (1994) 351.

\bibitem{Langanke04}
  K.\ Langanke, G.\ Mart{\'\i}nez-Pinedo, P.\ von Neumann-Cosel,
  and A.\ Richter,
  {\em nucl-th/0402001\/}.

\bibitem{Fujita96}
  Y.\ Fujita\doetal{, H.\ Akimune, I.\ Daito, M.\ Fujiwara, 
    M.N.\ Harakeh, T.\ Inomata, J.\ J{\"a}necke, K.\ Katoria, 
    H.\ Nakada, S.\ Nakayama, A.\ Tamii, M.\ Tanaka, H.\ Toyokawa and 
    M.\ Yosoi}, {\em Phys.\ Lett.\ \/} {\bf B 365} (1996) 29.

\bibitem{SNS}
  F.T.\ Avignone III, L.\ Chatterjee, Y.V.\ Efremenko, and M.R.\ Strayer, 
  {\it J.\ Phys.\ \/} {\bf G 29} (2003) 2497.

\bibitem{Langanke00}
  K.\ Langanke and G.\ Mart{\'\i}nez-Pinedo,
  \BibTitle{Shell-model calculations of stellar weak interaction
  rates: II.\ Weak rates for nuclei in the mass range $A=45-65$ in
  supernovae environments,}
  {\it Nucl.\ Phys.\ \/} {\bf A 673} (2000) 481.

\bibitem{Martinez96}
   G.\ Mart{\'\i}nez-Pinedo,
     A.\ Poves, E.\ Caurier, and A.P.\ Zuker,
   \BibTitle{Effective $g_A$ in the $pf$ shell,}
   {\it Phys.\ Rev.\ \/} {\bf C 53} (1996) R2602.

\bibitem{gagv}
  I.S.\ Towner and J.C.\ Hardy, in: W.C.\ Haxton, E.M.\ Henley (Eds.),
  Symmetries and Fundamental Interactions in Nuclei,
  World Scientific, Singapore, 1995, p.\ 183.

\bibitem{exp-spectrum} 
  Online version of Nuclear Data Sheets: {\em
  http://ie.lbl.gov/ensdf/welcome.htm\/}

\bibitem{Kolbe99}
  E.\ Kolbe, K.\ Langanke, and P.\ Vogel, 
  \BibTitle{Weak reactions on \up12C within the continuum random phase
  approximation with partial occupancies,}
  {\it Nucl.\ Phys.\ \/} {\bf A 652} (1999) 91.

\bibitem{Antoine} 
 E.\ Caurier, computer code ANTOINE, IReS, Strasbourg, 1989.

\bibitem{Sampaio01}
 J.M.\ Sampaio, K.\ Langanke, and G.\ Mart\'{\i}nez-Pinedo,
 \BibTitle{Neutrino absorption cross sections in the supernova environment,}
 {\it Phys.\ Lett.\ \/} {\bf B 511} (2001) 11.

\bibitem{Heger01}
  A.\ Heger,
  K.\ Langanke, G.\ Mart{\'\i}nez-Pinedo, and S.E.\ Woosley,
  \BibTitle{Presupernova collapse models with improved
  weak-interaction rates,}
  {\it Phys.\ Rev.\ Lett.\ \/} {\bf 86} (2001) 1678.

\bibitem{Lattimer91}
  J.M.\ Lattimer and F.D.\ Swesty,
  {\it Nucl.\ Phys.\ \/} {\bf A 535} (1991) 331.

\bibitem{FFN82}
  G.M.\ Fuller, W.A.\ Fowler, and M.J.\ Newman, 
  \BibTitle{Stellar weak interaction rates for intermediate-mass
  nuclei, II. $A$=21 to $A$=60,}
  {\it Astrophys.\ J.\ \/} {\bf 252} (1982) 715.

  G.M.\ Fuller, W.A.\ Fowler, and M.J.\ Newman,
  \BibTitle{Stellar weak interaction rates for intermediate-mass
  nuclei, III. Rate tables for the free nucleons and nuclei with
  $A$=21 to $A$=60,}
  {\it Astrophys.\ J.\ Suppl.\ \/} {\bf 48} (1982) 279.

\bibitem{Gazit04}
  D.\ Gazit and N.\ Barnea,
  \BibTitle{Neutrino neutral reaction on $^4$He, effects of final
  state interaction and realistic NN force,}
  {\em nucl-th/0402077\/}.

\end{thebibliography}
\end{document}